\documentclass[prd]{revtex4}
\usepackage{epsfig} 
\usepackage{bm} 
\usepackage{amssymb} 
\usepackage[fleqn]{amsmath} 
\usepackage{amsfonts} 
\usepackage{graphicx} 
\usepackage{psfrag} 
\usepackage{bm} 
\usepackage{setspace}
\def\bdm{\begin{displaymath}} 
\def\edm{\end{displaymath}} 
\def\be{\begin{equation}} 
\def\ee{\end{equation}} 
\def\ba{\begin{array}} 
\def\ea{\end{array}}

\begin{document}
\title {Testing the accuracy of the overlap criterion} 
\author{M.~Mestre}\email{mmestre@fcaglp.unlp.edu.ar} 
\author{P.~M.~Cincotta}\email{pmc@fcaglp.unlp.edu.ar} 
\author{C.~M.~Giordano}\email{giordano@fcaglp.unlp.edu.ar}

\affiliation{Facultad de Ciencias Astron\'omicas y Geof\'{\i}sicas, Universidad 
Nacional de La Plata and Instituto de Astrof\'{\i}sica de La Plata (CONICET), 
Paseo del Bosque, B1900FWA La Plata, Argentina}

\begin{abstract} 
Here we investigate the accuracy of the overlap criterion 
when applied to a simple near--integrable model in both its 2D and 3D version. 
To this end, we consider respectively, two and three 
quartic oscilators as the unperturbed system, and couple the degrees of freedom by a cubic, 
non--integrable perturbation.  
For both systems we compute the unperturbed resonances up to order  
${\mathcal O}(\epsilon^2)$, and model each 
resonance by means of the pendulum approximation in order to 
 estimate the theoretical critical value of the perturbation parameter 
for a global transition to chaos. 
We perform several surface of sections for the bidimensional case to derive 
an empirical value to be compared to our theoretical estimation, 
being both in good agreement. 
Also for the 3D case a numerical estimate is attained 
that we observe matches the critical value resulting from theoretical means.   
This confirms once again that reckoning resonances up to  
${\mathcal O}(\epsilon^2)$  suffices in order the overlap criterion to work out. 
  
\end{abstract} 
 
\maketitle 

\section{Introduction} 
 
Though the stability problem of Hamiltonian systems has been almost
completely elucidated by a rigourous sequence of theorems that build up
the so called KAM theory (see for instance \cite{CH79} and \cite{Re}, together 
with the original references therein: \cite{Kolmogorov}, \cite{Moser} and \cite{Arnold}), 
the application of the results of the KAM theory to a specific
system is far from being an easy task. In fact, it turns out to be much
simpler to take advantage of the heuristic Overlap Criterion, which seems
to provide similar estimations to those resulting from the KAM theory.

The overlap criterion due to Chirikov (see
\cite{CH79}) has been largely used in many different fields,  
its  probably most popular application being the study of inestabilities in the 
Solar System as well as in other planetary models 
(see for example \cite{W80}, \cite{L95}, \cite{M06}).  
In any case, since the widespread  model for a resonance is the pendulum approximation,  
the overlap criterion applies directly to the intersection of their 
associated unperturbed separatrices (or heteroclinic intersections). 

In his pioneer work on the standard map~\cite{CH79} Chirikov  
shows that the application of the overlap criterion to primary resonances 
overestimates the actual value of the critical parameter, $K_{\mathrm{c}}$,
and only when high order resonances are considered, does the overlap criterion 
succeed in providing a more accurate value for $K_{\mathrm{c}}$. In fact, the author shows that on including 
the third harmonics resonances, the overlap criterion leads to $K_{\mathrm{c}}\approx 1$, rather
close to the empirical value.
 
In the present effort we address a similar analysis to that performed by Chirikov, 
but using a 2D and a 3D near--integrable  
Hamiltonian systems, namely, two and three uncouppled quartic oscillators perturbed by a cubic 
term.  The  3D version of this model has been studied  
in \cite{CGS03}, \cite{GC04}, where the authors numerically investigate the global dynamical 
properties of the model and estimate the 
critical value $\epsilon_{\mathrm{c}}$ beyond which the system is globally chaotic, i.e. 
for which less than the 10\% of the energy surface corresponds to invariant tori. 
By means of the overlap criterion we derive such a critical value for the perturbative 
parameter  $\epsilon$ on considering not only primary but also high order resonances, 
and the perturbation Fourier series truncated at ${\mathcal O}(23^{-2})$ in their   
coefficients. 
We then compare, for each case, the theoretical critical value with that obtained by numerical means.

The paper is organized as follows. The dynamical system under study in its 2D version is described in Section II    
 and its relevant resonances at $\mathcal{O}(\epsilon)$ are obtained in Section III, their 
 widths being determined in Section IV.
 The resonances at $\mathcal{O}(\epsilon^2)$ are provided in Section V, where an estimate of the critical value 
 of the perturbative parameter is provided.  
 For the sake  of comparison, an empirical estimate of such a value is obtained in 
  Section VI by recourse of performing several surfaces of section for the system.
 The 3D model is addressed in Section VII, whose resonances at order $\mathcal{O}(\epsilon)$  
 and $\mathcal{O}(\epsilon^2)$ are given in Sections VIII and IX respectively.
 Section X is devoted to the theoretical estimate of the critical parameter, which is shown to be in good agreement  
 with the one given in \cite{GC04}. A final discussion is provided in Section XI.

\section{The 2D dynamical system} 
 
Here we will be concerned first with a 2D perturbed quartic oscillator.  
In cartesian coordinates the system is described by the following Hamiltonian 
(see \cite{CGS03}): 
 
\be 
\label{pertquartic} 
\tilde{H}(\bm{p},\bm{q})=\frac{1}{2}(p_x^2+p_y^2)+\frac{1}{4}(x^4+y^4)+\epsilon x^2y,  
\ee 
 
\noindent  
where $\epsilon$ is a perturbative parameter that controls the strength of the  
perturbation. On setting $\epsilon=0$ we recover the integrable quartic 
oscillator's Hamiltonian   
(see~\cite{GIR},\cite{CGS03} and references therein), whose  
solutions are given by  
 
\be 
\label{quarticsolution} 
\ba{l} 
x(t)=x_0(h_1) \; \sum_{n=1}^\infty \alpha_n \cos\big((2n-1)\omega_1(h_1)t\big),\\ \\ 
y(t)=y_0(h_2) \; \sum_{n=1}^\infty \alpha_n \cos\big((2n-1)\omega_2(h_2)t\big), 
\ea 
\ee 
where we have used the following definitions:

\begin{equation} 
  \label{quarticdefinitions} 
  \begin{split}
    x_0(h_1)&=4\beta {h_1}^{1/4},\\[3.5mm] 
    y_0(h_2)&=4\beta {h_2}^{1/4},\\[3.5mm] 
    \omega_i(h_i)&=\sqrt{2} \beta {h_i}^{1/4},\quad  i=1,2\\[3.5mm]  
    \alpha_n&=\frac{1}{\cosh\big((n-1/2)\pi\big)},\\[3.5mm] 
     \beta&=\pi/2K(1/\sqrt{2})\approx 0.847, 
    \end{split}
\end{equation} 
where $K(k)$ denotes the complete elliptic integral, 
and the coeficients in the Fourier expansions (\ref{quarticsolution}) satisfy:
$$\frac{\alpha_{n+1}}{\alpha_n} \approx \frac{1}{23}.$$ 

The third equation in (\ref{quarticdefinitions})  reveals the dependence of the frequencies on the unperturbed  
energies, enabling us to get the functional relationship between the latter and  
the unperturbed action variables, namely, $h_i=A{I_i}^{4/3}$, with $A=(3\beta/2\sqrt{2})^{4/3}$. 
 
With this relation in mind, and taking into account that the angle variables   
 are  $\theta_i\equiv\omega_i(h_i)t, \, i=1,2$, the  
complete Hamiltonian, in terms of the action-angle variables of 
the unperturbed Hamiltonian, can be recast as: 
 
\begin{equation} 
\label{Hamiltoniano} 
H(\bm{I},\bm{\theta}) = H_0(\bm{I})+\epsilon V(\bm{I},\bm{\theta}), 
\end{equation} 
where
 
\begin{eqnarray} 
H_0(\bm{I}) & = &  A({I_1}^{4/3}+I_2^{4/3}),\nonumber\\ 
  V(\bm{I},\bm{\theta})& = &   
  \hat{V}(\bm{I}) \sum_{n,m,k=1}^{\infty}\alpha_{nmk} 
  \left\{\cos\big(2(n+m-1)\theta_1\pm(2k-1)\theta_2 \big)+\cos\big( 2(n-m)\theta_1\pm(2k-1)
\theta_2 \big)\right\} 
 \label{Hamiltoniano1} 
\end{eqnarray} 
with $\alpha_{nmk}\equiv \alpha_{n}\alpha_{m}\alpha_{k}$, and   
$\hat{V}(\bm{I})\equiv 2^{5/2}3\beta^4 I_1^{2/3} I_2^{1/3}$, the $\pm$  
sign meaning that both terms are included in the series. 
 
\section{Resonances at $\mathcal{O}(\epsilon)$} 
 
A glance at the perturbation series given in equation (\ref{Hamiltoniano1})  
reveals that the number of resonant terms at first order  
in the perturbative parameter is unbounded, which is a drawback to take into account 
the width of every resonance at such an order. 
 
However, the strong dependence of the Fourier amplitudes on $(n+m+k)$, through  
the quantities $\alpha_{nmk}\approx 1/23^{(n+m+k-3)}$, gives us a good 
hint on how to gather the $\mathcal{O}(\epsilon)$ resonances and where to  
cut the series, our attempt being to keep terms only up to $\mathcal{O}(1/23^2)$.  
 
All the possible combinations of $n$, $m$ and $k$ verifying   
that $n+m+k \leq 5$, yield 24 different vectors which are listed in  
Table \ref{vectores1}, together with the number of times they appear in a   
term with coefficient $\alpha_{nmk}$ of a given order in 1/23.  
Thus, $N_0$ denotes the number of times the vector appears with coefficient 
$\alpha_{nmk}=\alpha_1^3$, $N_1$ the number of times it arises with  
coefficient $\alpha_1^2\alpha_2(\approx \alpha_1^3/23)$, and $N_2$  
corresponds to either the coefficient $\alpha_1\alpha_2^2$ or $\alpha_1^2\alpha_3$ 
(which are approximately $\alpha_1^3/23^2$). From now on, these vectors    
will be appointed as  harmonics at order $\mathcal{O}(\epsilon,1/23^2)$. 
 
\begin{table}[!ht] 
\centering 
\begin{tabular}{|c|c|c|c|c|c|c|c|} \hline 
 
 vector   & $N_0$ & $N_1$ & $N_2$ & vector  & $N_0$ & $N_1$ & $N_2$\\ \hline 
(2,1) & 1 &   1  &   0  &   (2,-1) & 1 &   1  &   0  \\ 
(0,1) & 1 &   0  &   1  &    (0,-1) & 1 &   0  &   1  \\ 
(2,3) & 0 &  1 & 1 & (2,-3) & 0 &   1  &   1  \\ 
(0,3)& 0 & 1&0 &    (0,-3) & 0 &   1  &   0  \\ 
(2,5)&0&0&1 &     (2,-5) & 0 &   0  &   1  \\ 
(0,5)&0&0&1 &    (0,-5) & 0 &   0  &   1  \\ 
(4,1)&0&2&1 &    (4,-1) & 0 &   2  &   1  \\ 
(-2,1)&0&1&0 &   (-2,-1) & 0 &   1  &   0  \\ 
(4,3)&0&0&2&    (4,-3) & 0 &   0  &   2  \\ 
(-2,3)&0&0&1&    (-2,-3) & 0 &   0  &   1  \\ 
(6,1)&0&0&3&    (6,-1) & 0 &   0  &   3  \\ 
(-4,1)&0&0&1&    (-4,-1) & 0 &   0  &   1  \\\hline 
 
\end{tabular} 
\caption{Harmonics in the Fourier expansion (\ref{Hamiltoniano1}) at 
$\mathcal{O}(\epsilon,1/23^2).$} 
\label{vectores1} 
\end{table} 
 
On applying the resonance condition $\bm{m}\cdot\bm{\omega}=0$, with  
$\bm{m}\in \mathbb{Z}^2/\{\bm{0}\}$, to the unperturbed system, the following relation  
between the energies in each degree of freedom is obtained:  
\be 
\label{condicionresonancia} 
m_1 h_1^{1/4}+m_2 h_2^{1/4}=0,  
\ee 
which implies that the resonance structure in energy and action space  
consists of straight lines (with positive slope) given by   
\be 
\label{relacionentreenergias} 
h_2^r=\frac{m_1^4}{m_2^4}h_1^r, 
\ee
and 
\be  I_2^r=\left|\frac{m_1^3}{m_2^3}\right|I_1^r, 
\ee 
respectively.
 
Moreover, eq. (\ref{condicionresonancia}) indicates that $m_1 m_2\leq 0$, whence,   
 those vectors having both components with the same sign must be discarded. 
Let us notice however, that not all of the remaining harmonics at order $\mathcal{O}(\epsilon,1/23^2)$ are 
actually resonant, as it will be discussed in the forthcoming section. 
 
\section{Width of the resonances at $\mathcal{O}(\epsilon)$ }   
 
On computing the widths of resonances, the pendulum's approximation (see  
for instance \cite{CH79}, \cite{C02})  
provides a suitable description whenever each resonance is assumed to be  
isolated from the rest. 
 
Notice should be taken that, before proceeding to estimate the width of a given resonance, 
all the coefficients $\alpha_{nmk}$  associated to the same trigonometric function are to be added together into a single one.  
Indeed, for each given vector $\bm{m}$, we should define the coefficient: 
\bdm 
\alpha_{\bm{m}}\equiv \sum_{m+n+k\leq 5}\alpha_{nmk}, 
\edm 
where $n$, $m$, and $k$ are natural numbers that  
combine to form the vector $\bm{m}$ in any of the four ways displayed in  
eq. (\ref{Hamiltoniano1}). 
Let us remark  that  
$\alpha_{\bm{m}}\approx \alpha_1^3(N_0+N_1/23+N_2/23^2)$. 
 
Inasmuch the pendulum's approximation has been applied to this single resonant  
term, the new (resonant) Hamiltonian turns out to be:    
 
\be 
\label{resonantHamiltonian1} 
H_r(p_1,\psi_1)=\frac{p_1^2}{2M}+\epsilon \hat{V}(\bm{I^r})\alpha_{\bm{m}}\cos\psi_1, 
\ee 
with 
\be 
M^{-1}\equiv m_i\frac{\partial\omega_i^r}{\partial I_j}m_j=m_i^2\frac{\partial\omega_i^r}{\partial I_i},\qquad 
\bm{I}=\bm{I^r}+\bm{m}p_1,\qquad \psi_1=\bm{m}\cdot\bm{\theta}, 
\ee 
where the sum over repeated indexes should be understood.
 
Let $p_r$ be the maximum variation of $p_1$ within the oscillation regime, then  
\be 
\label{pr} 
p_r= 2 \left(\epsilon \,M\hat{V}(\bm{I^r})\alpha_{nmk}\,\right)^{1/2}=\,2^{7/2}\beta^{1/2} \;{\left|\frac{m_1^3 m_2^{-1}}{m_1^4+ m_2^4}\right|}^{1/2}\epsilon^{1/2}\,\alpha_{nmk}^{1/2}\;(h_1^r)^{5/8}. 
\ee 
 
As a consequence of the simple pendulum dynamics, the  
maximum displacement of the unperturbed action variables  
depends on $\bm{m}$ and $p_r$ in the fashion: 
${(\Delta \bm{I})}^{\bm{r}}\equiv {(\bm{I}-\bm{I^r})}_{max} =p_r \bm{m}$. 
 
Furthermore, the maximum displacement of the unperturbed energy is 
given by  
$\left|{(\Delta h_i)}_{\bm{m}}^r \right|= 
\left|\omega_i^r (\Delta I_i)_{\bm{m}}^r\right|$;  
and it is the maximum amplitude  
attained in the oscillation of any of the unperturbed energies 
that measures the width of the resonance. 

Let us now recall that in any 2D problem the energy conservation  
condition $h=h_1+h_2$, together with the resonance condition  
$m_1h_1^{1/4}+m_2h_2^{1/4}=0$ allow both  
$h_1^r$ and $h_2^r$ to be written in terms of the total unperturbed energy $h$. 
Thus, the amplitude can be recast in terms of  
$m_1$, $m_2$, $\epsilon$, $\alpha_{\bm{m}}$ and $h$ as follows: 
 
\begin{equation} 
\label{anchoresonancia1} 
\begin{split}
    |(\Delta h_1)_{\bm{m}}^r| &= 2^4\beta^{3/2}\;\frac{|m_1|^{5/2}|m_2|^3}{|m_1^4+m_2^4|^{11/8}} 
\;\epsilon^{1/2} \,\alpha_{\bm{m}}^{1/2} \; h^{r7/8},\\[3.5mm] 
|(\Delta h_2)_{\bm{m}}^r| &= |(\Delta h_1)_{\bm{m}}^r|.
\end{split}
\end{equation}

The last identity in (\ref{anchoresonancia1}) is due to the fact that in  
presence of a single resonance, the motion of the system is tangent to  
the unperturbed energy surface, and for this particular model such a 
surface is given by $h=h_1+h_2$, which leads to $\Delta h_2=-\Delta h_1$.  

\begin{figure}[!ht]
  \centering
  \includegraphics[angle=-90,width=.50\textwidth]{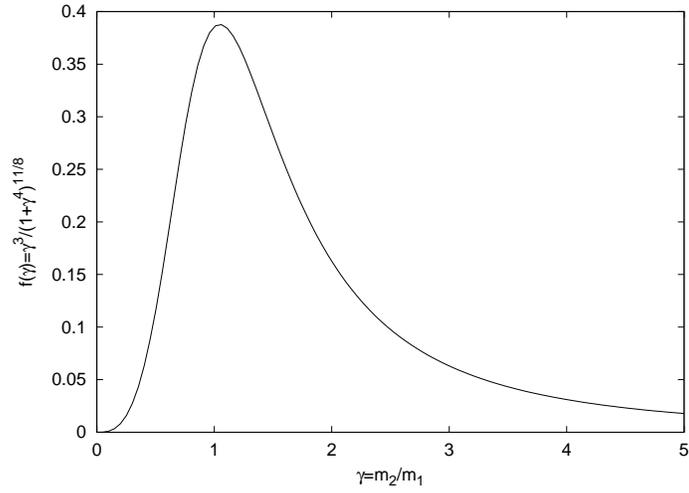}
  \caption{Dependence of the resonance width on $m_2/m_1$.}
  \label{fig:depcongamma}
\end{figure}

We note that the width of the resonances at $\mathcal{O}(\epsilon,1/23^2)$ 
depends on the harmonic numbers in the manner shown in 
Fig.~\ref{fig:depcongamma}, so that those resonances with values of 
$\gamma=m_2/m_1$ out of the range $[0.5,2.5]$ should be narrow. 

A glance at both eq. (\ref{anchoresonancia1}) and Fig.~\ref{fig:depcongamma} reveals that those harmonic vectors  
with any of its components equal to zero do not change the energies, and 
consequently, they should not be considered as resonant vectors.
 
Further, since the perturbation terms are even, if $\bm{m}$ is a  
resonant vector then $-\bm{m}$ is also a resonant one (both  
corresponding to the same resonance). Hence, for each resonance    
just one representative resonant vector can be considered, encompassing  
into its concomitant coefficient the contribution corresponding to 
its opposite vector as well. 
All the relevant data required to compute the width of each resonance at   
$\mathcal{O}(\epsilon,1/23^2)$ is displayed in Table \ref{vectores1bis}, 
where we have also included the value  
of $h_1^r$ corresponding to a total unperturbed energy of $h=1/(4\beta^4)\approx 0.485$ (the one used in \cite{CGS03}). 
  
\begin{table}[!ht] 
\centering 
\begin{tabular}{|c|c|c|c|c|} \hline 
 
vector   & $N_0$ & $N_1$ & $N_2$ &  $h_1^r$  \\ \hline 
(2,-5) & 0 & 0 & 1 & 0.4729 \\ 
(2,-3) & 0 & 1 & 2 & 0.4050 \\ 
(4,-3) & 0 & 0 & 2 & 0.1166 \\ 
(2,-1) & 1 & 2 & 0 & 0.0285 \\ 
(4,-1) & 0 & 2 & 2 & 0.0019 \\ 
(6,-1) & 0 & 0 & 3 & 0.0004 \\ \hline 
 
\end{tabular} 
\caption{Resonant vectors at $\mathcal{O}(\epsilon)$ up to $\mathcal{O}(1/23^2)$.} 
\label{vectores1bis} 
\end{table} 
 
\begin{figure}[!ht] 
  \centering 
  \includegraphics[width=.60\textwidth]{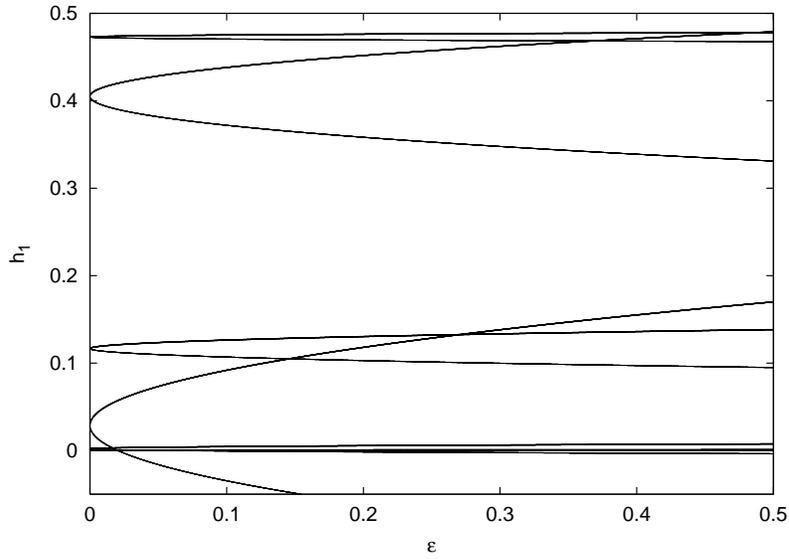} 
  \caption{Widths of those resonances at $\mathcal{O}(\epsilon)$ up to $\mathcal{O}(1/23^2)$, in terms of $\epsilon$.} 
  \label{fig:anchos} 
\end{figure} 
 
We have computed the resonance widths corresponding to $\epsilon$ in the  
range $[0,0.5]$.  
Fig. \ref{fig:anchos} displays both the maximum and minimum   
values of $h_1$ for each resonance vs. the perturbative parameter 
$\epsilon$;  
the total unperturbed energy being $h=1/(4\beta^4)$. 
We observe that for $\epsilon\sim 0.15$, the   
$(6,-1), (4,-1), (2-1)$  and $(4,-3)$  resonances do overlap, but lie far away from the $(2,-3)$ and $(2,-5)$ resonances. Therefrom we could infer 
that the energy surface presents two unconnected regions of chaotic motion,  so that a global transition to chaos does not take place  for  
$\epsilon\lesssim 0.5$, which leads to a critical theorical 
value for the perturbation parameter $\epsilon_{\mathrm{c}}\gg 0.5$.

\section{Resonances at $\mathcal{O}(\epsilon^2)$} 
 
In those regions of phase space which are far from any   
primary resonance (i.e. where the diophantic condition holds 
for every primary resonance), we can introduce new cononical variables, 
$(\bm{J},\bm{\varphi})$, in such a fashion that the transformed 
Hamiltonian  
consists of a part depending on the new momentum and a  
perturbation that, though being non--integrable, has an  
amplitude of $\mathcal{O}(\epsilon^2)$. 
 
\begin{eqnarray} 
\mathcal{H}(\bm{J},\bm{\varphi})= H_0(\bm{J})+\epsilon^2 \left\{\frac{1}{2}\frac{\partial^2 H_0(\bm{J})}{\partial J_j \partial J_i} \frac{\partial \Phi(\bm{J},\bm{\varphi})}{\partial \varphi_j}\frac{\partial \Phi(\bm{J},\bm{\varphi})}{\partial \varphi_i} 
+ \frac{\partial V(\bm{J},\bm{\varphi})}{\partial J_i}\frac{\partial \Phi(\bm{J},\bm{\varphi})}{\partial \varphi_i}\right\}+\mathcal{O}(\epsilon^3) 
\label{Hamiltoniano2}, 
\end{eqnarray} 
where $\Phi$ stands for the trigonometric part 
of the generatrix function of the canonical 
transformation: 
 
\begin{equation}  
F(\bm{J},\bm{\theta})= \bm{J}\cdot\bm{\theta}+\epsilon\,
\Phi(\bm{J},\bm{\theta})\qquad
\Phi(\bm{J},\bm{\theta})=\sum_{\bm{m} \in \mathbb{Z}^2/\{\bm{0}\}}\Phi_{\bm{m}}(\bm{J})\sin (\bm{m}\cdot\bm{\theta})\nonumber, 
\end{equation} 
with $\Phi_{\bm{m}}(\bm{J})=-V_{\bm{m}}(\bm{J})/\bm{m}\cdot\bm{\omega}(\bm{J})$. 
 
After computing the right side of expression (\ref{Hamiltoniano2}) one 
finds: 
\begin{eqnarray} 
\mathcal{H}(\bm{J},\bm{\varphi}) = H_0(\bm{J})\;+\;\epsilon^2 \sum_{\bm{m},\bm{m'}}\mathbb{C}(\bm{m},\bm{m'},\bm{J})
\left\{\cos\big((\bm{m}+\bm{m'})\cdot\bm{\varphi}\big)+\cos\left((\bm{m}-\bm{m'})\cdot\bm{\varphi}\right)\right\} \;+\; \mathcal{O}(\epsilon^3), 
\label{Hamiltoniano3} 
\end{eqnarray} 
where the coefficients are given by  
\begin{equation*} 
\mathbb{C}(\bm{m},\bm{m'},\bm{J})=\alpha_{\bm{m}}\alpha_{\bm{m'}} \left\{2^3 3^{4/3}\beta^{28/3} \frac{(m_1m'_1J_1^{2/3}J_2^{2/3}+m_2m'_2J_1^{4/3})}{(\bm{m}\cdot\bm{\omega}(\bm{J}))(\bm{m'}\cdot\bm{\omega}(\bm{J}))} 
-2^43\beta^8 \frac{(2m'_1J_1^{1/3}J_2^{2/3}+m'_2J_1^{4/3}J_2^{-1/3})}{(\bm{m'}\cdot\bm{\omega}(\bm{J}))}\right\}. 
\end{equation*}  
 
There are several relevant  facts to be remarked: 
(i)~on working at $\mathcal{O}(\epsilon)$ we have only considered resonances  
 up to $\mathcal{O}(1/23^2)$ in the Fourier coefficients; therefore, the  
series in equation (\ref{Hamiltoniano3}) should actually be replaced  
by a finite sum over those harmonics, $\bm{m}$ and $\bm{m'}$,  
whose associated coefficients ($\alpha_{\bm{m}}$ and  
$\alpha_{\bm{m'}}$, respectively) are such that their product is of 
order either $\alpha_1^6$, $ \alpha_1^6/23$ or $\alpha_1^6/23^2$; 
(ii)~there are many different pairs of harmonics $(\bm{m},\bm{m'})$ 
at $\mathcal{O}(\epsilon)$,   
 which combine into the same  harmonic $\bm{n}$ at $\mathcal{O}(\epsilon^2)$; 
(iii)~as a consequence of the evenness of the perturbation term, if    
$\bm{n}$ is a resonant vector $-\bm{n}$ is also a resonant one; 
(iv)~the resonance condition implies that $n_1n_2\leq 0$; and 
(v)~the condition of being far from resonances at   
$\mathcal{O}(\epsilon)$ implies that we must discard all those harmonics  
which are a multiple of any resonant vector at $\mathcal{O}(\epsilon,1/23^2)$. 

To cope with the situation set up by the issues (ii) and (iii), we  have added  all the concomitant contributions into a single coefficient $\mathbb{D}$, namely,  
\begin{equation} 
\mathbb{D}(\bm{n},\bm{J})=\sum_{\bm{m},\bm{m'}}\mathbb{C}(\bm{m},\bm{m'},\bm{J}), 
\end{equation} 
where the sum extents to all the harmonics $(\bm{m},\bm{m'})$ at $\mathcal{O}(\epsilon,1/23^2)$ such that $\bm{n}=\bm{m}+\bm{m'}$, $\bm{n}=\bm{m}-\bm{m'}$, $-\bm{n}=\bm{m}+\bm{m'}$,  or $-\bm{n}=\bm{m}-\bm{m'}$, and for which $\mathcal{O}(\alpha_{\bm{m}}\alpha_{\bm{m'}})$ is not greater than $\alpha_1^6/23^2$.  
 
Taking into account all the above mentioned considerations, the Hamiltonian can be written in the form: 
 
\begin{equation} 
\mathcal{H}(\bm{J},\bm{\varphi})=H_0(\bm{J})+\epsilon^2 \sum_{\bm{n}}\mathbb{D}(\bm{n},\bm{J})\cos(\bm{n}\cdot\bm{\varphi}). 
\label{Hamiltoniano4} 
\end{equation} 
 
Therefore, in the vicinity of a resonant torus $\bm{J^r}$, and by recourse of the  
pendulum approximation, we obtain the new resonant Hamiltonian:

\begin{equation} 
\label{resonantHamiltonian2} 
\mathcal{H}_r({\mathcal{P}}_1,\Psi_1)=\frac{{{\mathcal{P}}_1}^2}{2\mu}+\mathcal{U}_0 \cos(\Psi_1), 
\end{equation} 
where  
 
\begin{eqnarray} 
\mu^{-1}&\,\equiv& n_i\frac{\partial\omega_i^r}{\partial J_j}n_j, \nonumber\\ 
\mathcal{U}_0 \,&\equiv&\,\epsilon^2 \,\mathbb{D}(\bm{n},\bm{J^r}). 
\end{eqnarray} 
 
Thus, the maximum displacement of the unperturbed action variables is given by ${(\Delta \bm{J})}^{\bm{r}}\equiv {(\bm{J}-\bm{J^r})}_{max} =\mathcal{P}_r \bm{n}$, with $\mathcal{P}_r= 2 \left(\mu |\mathcal{U}_0|\right)^{1/2}$,  
and the widths of the resonances at $\mathcal{O}(\epsilon^2)$ 
are given by  
\begin{eqnarray} 
|(\Delta h_1)_{\bm{n}}^r| = 2^4\beta^3 h^{3/4}\epsilon\frac{|n_2|^3 n_1^2}{(n_1^4+n_2^4)^{5/4}}\cdot 
\left| \sum_{\bm{m},\bm{m'}} \alpha_{\bm{m}} \alpha_{\bm{m'}} \left\{ \frac{2(2|n_1|^3 m'_1+|n_2|^3 m'_2)}{|n_1n_2|(m'_1|n_2|+m'_2|n_1|)}\pm \frac{n_1^2m'_1m_1+n_2^2m'_2m_2}{(m'_1|n_2|+m'_2|n_1|)^2} \right\} \right|^{1/2}, \nonumber 
\end{eqnarray} 
where the plus sign corresponds to $\bm{n}=\bm{m}+\bm{m'}$ and the minus sign to $\bm{n}=\bm{m}-\bm{m'}$, $m_i$ ought to be written in terms of $m'_i$ and $n_i$. 
As a consequence of this last expression,  resonant vectors are compelled to have no null components. 
 
\begin{table}[!ht] 
\centering 
\begin{tabular}{|c|c|} \hline 
vector   &  $h_1^r$  \\ \hline 
(2,-6) & 0.4791  \\ 
(2,-4) & 0.4565  \\ 
(2,-2) & 0.2425  \\ 
(6,-4) & 0.0800  \\ 
(6,-2) & 0.0059  \\ \hline 
\end{tabular} 
\caption{Resonant vectors of $\mathcal{O}(\epsilon^2)$ up to $\mathcal{O}(1/23^2)$.} 
\label{vectores2} 
\end{table} 
 
\begin{figure}[h!] 
  \centering 
  \includegraphics[width=.60\textwidth]{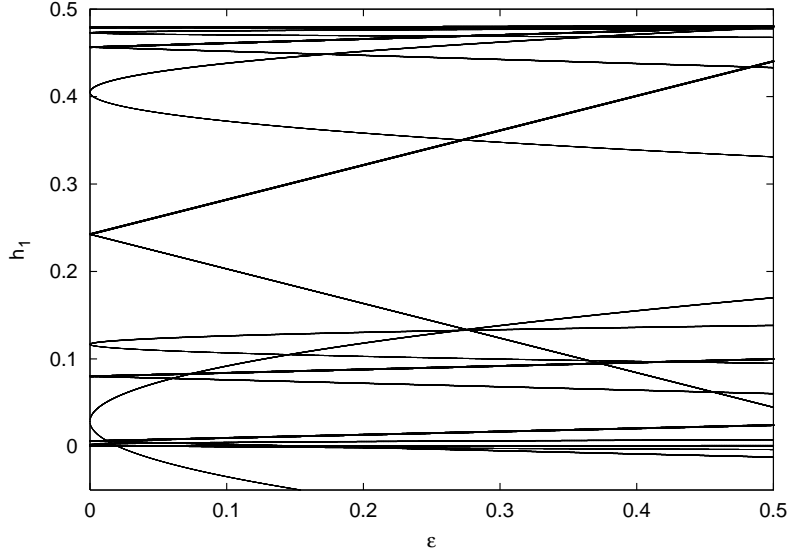} 
  \caption{Width of the resonances up to $\mathcal{O}(\epsilon^2,1/23^2)$ 
  vs. $\epsilon$.} 
  \label{fig:anchos2} 
\end{figure} 
 
The harmonics satisfying all the stated conditions for the perturbation at order $\epsilon^2$  
turn out to be just five, which are listed in Table~\ref{vectores2}. 
They will be referred to as resonant vectors at $\mathcal{O}(\epsilon^2,1/23^2)$. 
We have computed the concomitant resonance widths for $\epsilon$ in the range $[0,0.5]$. 
The results are presented in  Fig. \ref{fig:anchos2}, where also the resonances corresponding to $\mathcal{O}(\epsilon)$ have been included. 
Let us recall that the adopted value for the total unperturbed energy is $h=1/4\beta^4$. 
 
From Fig.~\ref{fig:anchos2} and Table~\ref{vectores2} we notice that the arising of the  
 $(2,-2)$ resonance connects the two  
sets of resonances that at $\mathcal{O}(\epsilon,1/23^2)$ appeared isolated for the considered 
$\epsilon$ range. The remaining resonances at $\mathcal{O}(\epsilon^2,1/23^2)$ appear completely 
overlapped with either set of resonances at $\mathcal{O}(\epsilon,1/23^2)$. From this plot, we could derive the critical value for the perturbative parameter, $\epsilon_{\mathrm{c}}\approx 0.28$.
 
\section{Numerical estimation of the critical value of the perturbation parameter} 
 
In this section we empirically estimate the value of $\epsilon_{\mathrm{c}}$ by means of Poincar\'e 
Surfaces of Section (SOS). To this aim, we take the intersections on 
the plane $x=0$ (actually $|x|<10^{-8}$) whenever $p_x>0$, for several initial conditions 
along the $y$--axis. 
   
\begin{figure}[!ht] 
   \begin{tabular}{cc} 
   \hspace{-4mm}\includegraphics[width=.5\textwidth]{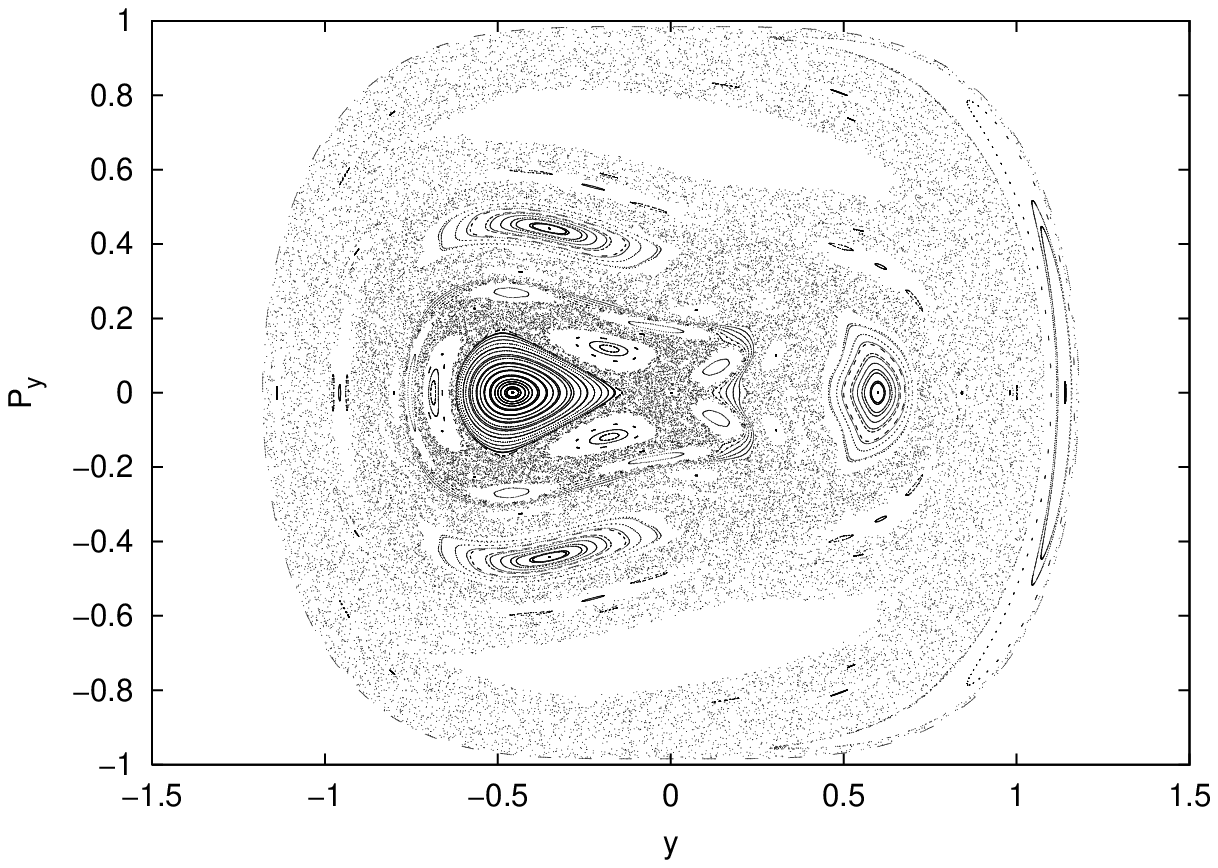} 
   \hspace{-4mm}\includegraphics[width=.5\textwidth]{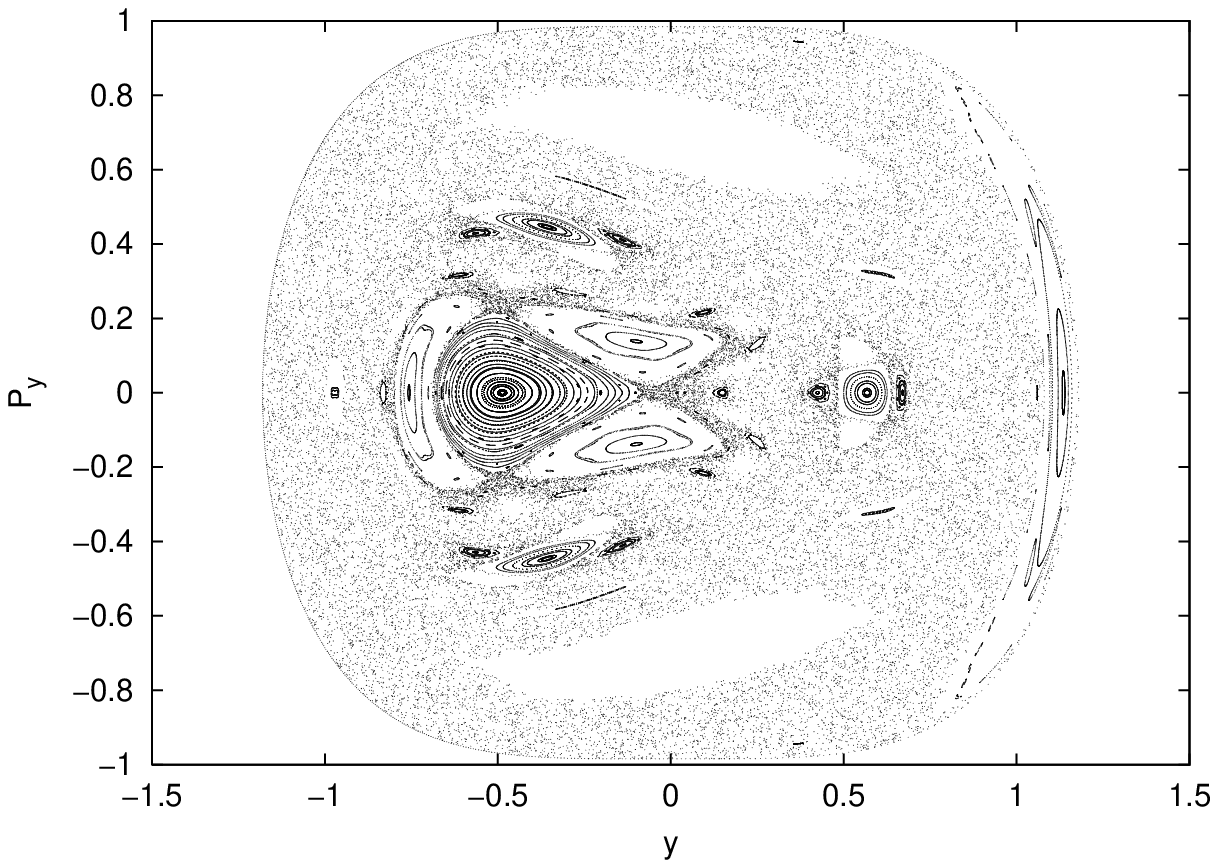} 
   \end{tabular} 
  \caption{Poincar\'e surfaces of section for $\epsilon=0.12$ (on the left), and for 
   $\epsilon=0.14$ (on the right).} 
  \label{fig:sec0.12} 
\end{figure}

Fig.\ref{fig:sec0.12} displays the SOS's corresponding to $\epsilon=0.12$~(on the left)  
and to $\epsilon=0.14$~(on the right), respectively. There  
we can distinguish the $(2,-1)$ resonance, very close to the last invariant 
curve that corresponds to the $y$--axis periodic orbit -- $(1,0)$ resonance--, and the $(2,-2), (2,-3)$ resonances as well.  
The $(2,-4)$, $(2,-5)$ and $(2,-6)$ resonances do not show up due to the fact that they are 
completely distroyed by overlap,  
as could be seen from Fig.~\ref{fig:anchos2} for this value of $\epsilon$. 
It is important to remark that several higher order resonances do appear which have not been  
theoretically computed.  

For $\epsilon=0.12$,  
we observe that the chaotic domain where the $(2,-4)$, $(2,-5)$ and $(2,-6)$ resonances appear distroyed by overlap is bounded by some KAM tori and thus, it remains unconnected with the outer chaotic component around the $(2,-3)$ and $(2,-1)$ resonances. 
On the other hand, for $\epsilon=0.14$, both chaotic zones are connected, leading to 
a global transition to chaos, in the sense that any orbit could explore almost 
all the chaotic component of phase space. 
 
\begin{figure}[!ht] 
  \centering 
  \includegraphics[width=.60\textwidth]{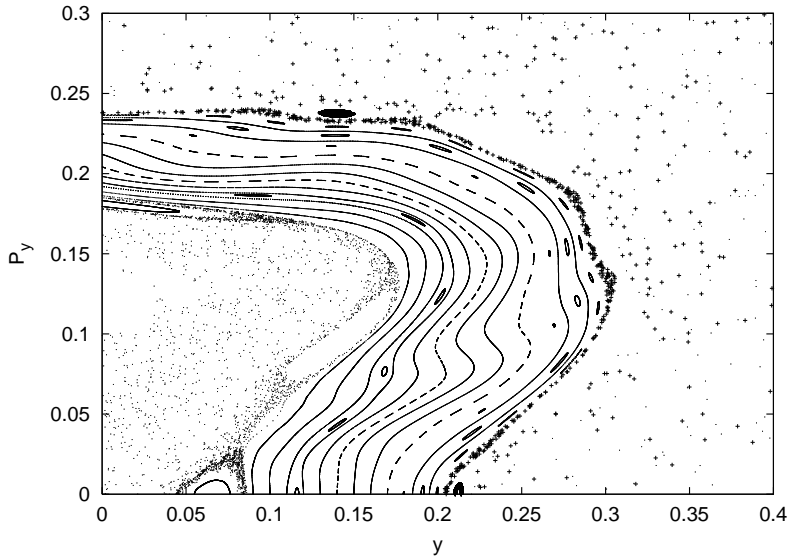} 
  \caption{Poincar\'e surface of section for $\epsilon=0.135$ illustrating the existence of KAM  
tori that separate both chaotic components.} 
  \label{fig:KAM} 
\end{figure}

Therefore, it looks like $\epsilon_{\mathrm{c}}$ lies somewhere in the range $(0.12,0.14)$. 
After performing a rather thorough numerical exploration, we have noted that for 
$\epsilon=0.135$, several KAM tori do persit, which are shown in 
Fig.\ref{fig:KAM}, where a zoom in the window $[0,0.4]\times [0,0.3]$ is presented. 
There, such KAM tori can be clearly distinguished and are seen to definitively separate 
both chaotic domains in phase space.  Nevertheless, this bounded region of chaotic
motion, does not involve the resonances we are taking into account to derive
the analitical esimation of $\epsilon_{\mathrm{c}}$, but high order ones.

Therefore, from experimental means, we may state that $\epsilon\gtrsim 0.135$ is  
a good lower bound for the critical value of the perturbation parameter.

\section{The 3D model}

Now we will focus on a 3D version of the dynamical system under study. 
In cartesian coordinates, its Hamiltonian is given by: 
 
\begin{equation} 
\label{3Dpertquartic} 
\tilde{H}(\bm{p},\bm{q})=\frac{1}{2}(p_x^2+p_y^2+p_z^2)+\frac{1}{4}(x^4+y^4+z^4)+\epsilon x^2(y+z).  
\end{equation} 
 
Let us notice that for a null value of the perturbative parameter we recover the three independent 
one dimensional quartic oscillators, whose solutions $x(t)$ and $y(t)$ are the ones   
given by equation (\ref{quarticsolution}), while $z(t)$ allows for the expression: 
 
\begin{equation} 
z(t)=z_0(h_3)\sum_{n=1}^\infty \alpha_n \cos\big((2n-1)\omega_3(h_3)t\big), 
\end{equation} 
 
\noindent  
where $z_0(h_3)=4\beta {h_3}^{1/4}$. 
 
In terms of the action-angle variables of the unperturbed Hamiltonian, 
the complete Hamiltonian (\ref{3Dpertquartic}) can be recast as: 
 
\begin{equation} 
\label{3DHamiltoniano} 
H(\bm{I},\bm{\theta}) = H_0(\bm{I})+\epsilon V(\bm{I},\bm{\theta}), 
\end{equation} 
with 
 
\begin{equation}
  \begin{split} 
    H_0(\bm{I})=&A \,({I_1}^{4/3}+I_2^{4/3}+I_3^{4/3}),\\ 
    V(\bm{I},\bm{\theta})=&\hat{V}_{12}(\bm{I}) \sum_{n,m,k=1}^{\infty}\alpha_{nmk}
    \left\{\cos\big( 2(n+m-1)\theta_1\pm(2k-1)\theta_2 \big)+\cos\big( 2(n-m)
      \theta_1\pm(2k-1)\theta_2 \big)\right\}+\\ 
    &+\hat{V}_{13}(\bm{I}) \sum_{n,m,k=1}^{\infty}\alpha_{nmk} 
    \left\{\cos\big( 2(n+m-1)\theta_1\pm(2k-1)\theta_3 \big)+\cos\big( 2(n-m)
      \theta_1\pm(2k-1)\theta_3 \big)\right\}, 
    \label{3DHamiltoniano1} 
  \end{split}
\end{equation}
where the new quantities $\hat{V}_{1j}(\bm{I})\equiv 2^{5/2}3\beta^4 I_1^{2/3} I_j^{1/3}$ have been introduced; the $\pm$ sign meaning that both terms are included in the series.

\section{Resonances at $\mathcal{O}(\epsilon)$} 
 
The perturbation given in eq. (\ref{3DHamiltoniano1}) shows that for each  
combination of $n, m$ and $k$ there result 8 harmonic vectors $\bm{m}$. 
Again, due to the even character of the perturbation, we take just one representative 
resonant vector, $\bm{m}$, whose coefficient $\alpha_{\bm{m}}$ also encompasses the contribution 
of its opposite vector 
$-\bm{m}$; and keep only those harmonics such that 
$\mathcal{O}(\alpha_{\bm{m}}) \leqslant \mathcal{O}(1/23^2)$.  
 
Since the harmonic vectors at $\mathcal{O}(\epsilon,1/23^2)$ can be splitted into two groups,
we denote by $\mathcal{Y}$  the subset of vectors whose third component is zero and 
by $\mathcal{Z}$  that of vectors having their second component equal to zero. 
Therefore, the perturbation can be written in the fashion: 

\begin{equation} 
V(\bm{I},\bm{\theta})=\hat{V}_{12}(\bm{I}) \sum_{\bm{m}\in\mathcal{Y}}\alpha_{\bm{m}}
\cos (\bm{m}\cdot\theta)+\hat{V}_{13}(\bm{I}) \sum_{\bm{m}\in\mathcal{Z}}\alpha_{\bm{m}}
\cos (\bm{m}\cdot\theta). 
\end{equation} 
 
On applying the resonance condition, $\bm{m}\cdot\bm{\omega}=0$, with 
$\bm{m}\in \mathbb{Z}^3/\{\bm{0}\}$, to the unperturbed system, 
the following relation between the actions in each degree of freedom is obtained: 
 
\begin{equation} 
\label{3Dcondicionresonancia} 
m_1 I_1^{1/3}+m_2 I_2^{1/3}+m_3 I_3^{1/3}=0, 
\end{equation} 
so that each resonant vector could not have all its three components of 
the same sign. Furthermore, whenever a resonant vector has two of its components equal to zero, 
we get a null amplitude for the perturbation term. 

Thus, we obtain twelve different resonant vectors, grouped in the following set:
\begin{equation*}
  \begin{split}
  \mathcal{V}_r(\epsilon,1/23^2)=\big\{&(2,-1,0),(2,-3,0),(2,-5,0),(4,-1,0),(4,-3,0),(6,-1,0),(6,-1,0),(2,0,-1),(2,0,-3),(2,0,-5),\\
&(4,0,-1),(4,0,-3),(6,0,-1)\big\}
  \end{split}
\end{equation*}

For those vectors $\bm{m}\in \mathcal{Y}$, the resonance condition together with the 
energy conservation condition define a curve in action space (not just a point as 
in the 2D case) given by: 
\begin{equation} 
\label{curvaY} 
\left\{ \begin{array}{l} 
I_2^r=-\left(\frac{m_1}{m_2}\right)^3 I_1^r, \\ 
I_3^r=\left\{\frac{h}{A}-\left(1+\frac{m_1^4}{m_2^4}\right)(I_1^r)^{4/3}\right\}^{3/4}, 
\end{array}\right. 
\end{equation} 
 
\noindent 
where $I_1^r \in [0,I_{max}^{\mathcal{Y}}]$ with 
$I_{max}^{\mathcal{Y}}=\left(\frac{h/A}{1+(m_1/m_2)^4}\right)^{3/4}$.

The concomitant curve for $\bm{m}\in \mathcal{Z}$ is given by 
 
\begin{equation} 
\label{curvaZ} 
\left\{ \begin{array}{l} 
I_2^r=\left\{\frac{h}{A}-\left(1+\frac{m_1^4}{m_3^4}\right)(I_1^r)^{4/3}\right\}^{3/4},\\ 
I_3^r=-\left(\frac{m_1}{m_3}\right)^3 I_1^r,  
\end{array}\right. 
\end{equation} 
 
\noindent 
with $I_1^r \in [0,I_{max}^{\mathcal{Z}}]$, the upper bound being 
$I_{max}^{\mathcal{Z}}=\left(\frac{h/A}{1+(m_1/m_3)^4}\right)^{3/4}$.
In energy surface both kind of resonances define straight lines.
On applying the pendulum approximation,  we obtain a similar resonant Hamiltonian to that 
given by equation (\ref{resonantHamiltonian1}), namely, 
\begin{equation} 
\label{3DresonantHamiltonian1} 
H_r(p_1,\psi_1)=\frac{p_1^2}{2M}+\epsilon V_{\bm{m}}\cos\psi_1, 
\end{equation} 
 
\noindent 
where $V_{\bm{m}}=\hat{V}_{12}(\bm{I^r})\alpha_{\bm{m}}$ for $\bm{m}\in \mathcal{Y}$, 
and $V_{\bm{m}}=\hat{V}_{13}(\bm{I^r})\alpha_{\bm{m}}$ for $\bm{m}\in \mathcal{Z}$. 
 
In energy space, the resonance widths in each degree of freedom, are adequately described by 
\begin{equation} 
\label{3Danchoresonancia1} 
(\Delta h_i)_{\bm{m}}^r=\omega_i(I_i^r)(\Delta I_i)_{\bm{m}}^r=\frac{4}{3}AI_i^{1/3}
2(\epsilon M V_{\bm{m}})^{1/2}m_i= \frac{8}{3}AI_i^{1/3}(\epsilon M V_{\bm{m}})^{1/2}m_i. 
\end{equation}  
 
Let us remark that, while in the 2D model the resonance and energy conservation conditions 
force the resonance width to depend on just one  
variable, either $h,$ $h_1^r$ or $I_1^r$, in the 3D model the resonance width is a 
function of two independent variables, which we have chosen to be $I_1^r$ and $h$.

\begin{figure}[h]
   \begin{tabular}{cc}
   \hspace{-4mm}\includegraphics[width=.5\textwidth]{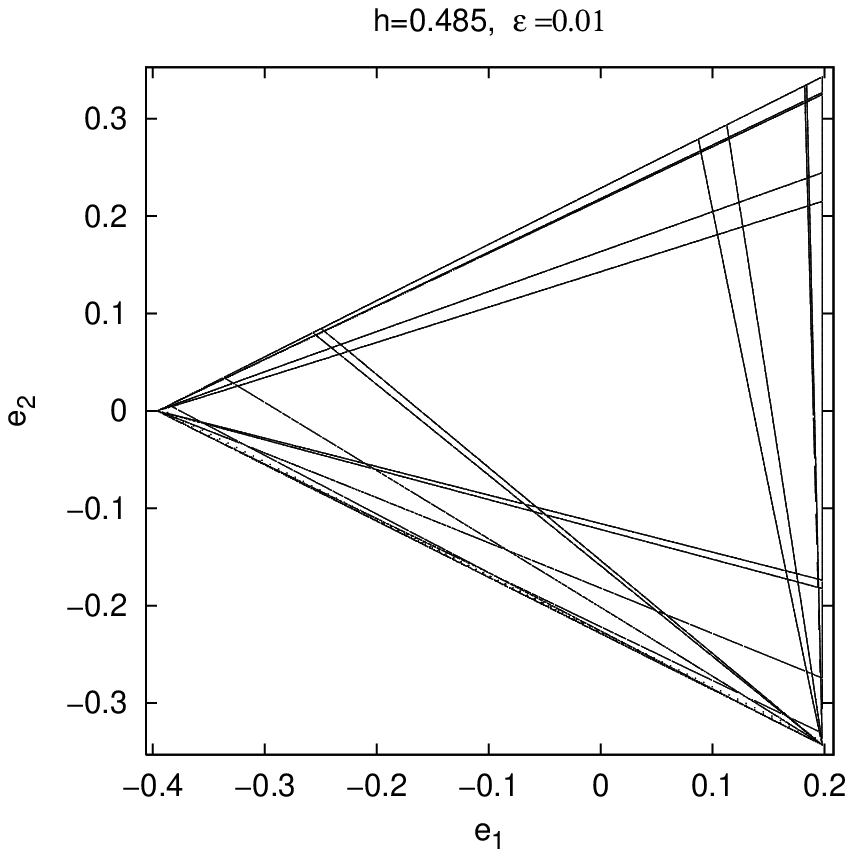}
   \hspace{-4mm}\includegraphics[width=.5\textwidth]{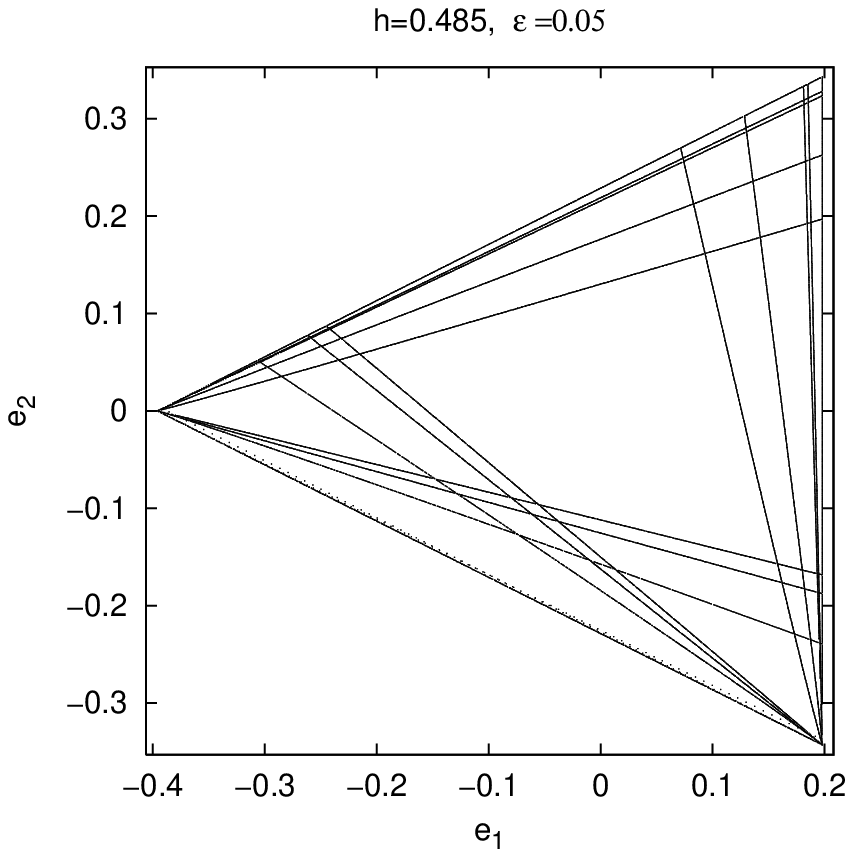}
   \end{tabular}
  \caption{Resonances at $\mathcal{O}(\epsilon, 1/23^2)$ for $\epsilon=0.01$ (on the left) and $\epsilon=0.05$ (on the right).}
  \label{3Dresonancias1}
\end{figure}

With the widths computed by means of (\ref{3Danchoresonancia1}), we can trace the 
displacements of the resonant energies, $h_i^r + (\Delta h_i)_{\bm{m}}^r$ and 
$h_i^r - (\Delta h_i)_{\bm{m}}^r$, for values of $I_1^r \in [0,I_{max}]$. 
Therefore, following \cite{CGS03} we perform the global change of coordinates:
\begin{eqnarray}  
\label{3Dtransformacion} 
e_1&=&\frac{1}{\sqrt{6}}(h_1-2h_2+h_3),\nonumber \\ 
e_2&=&\frac{1}{\sqrt{2}}(h_1-h_3), \nonumber \\ 
e_3&=&\frac{1}{\sqrt{3}}(h_1+h_2+h_3), 
\end{eqnarray} 
where $e_1 \in \left[-\sqrt{\frac{2}{3}}h,\frac{h}{\sqrt{6}}\right]$, 
$e_2 \in \left[-\frac{h}{\sqrt{2}},\frac{h}{\sqrt{2}}\right]$, $e_3=\frac{h}{\sqrt{3}}$, 
adopting the value  $h\approx 0.485$, to finally display in 
Fig.~\ref{3Dresonancias1} the region of energy surface occupied by the structure of resonances at $\mathcal{O}(\epsilon,1/23^2)$ for two different values of the perturbative parameter.
Let us  remark that many of the resonances in $\mathcal{V}_r(\epsilon,1/23^2)$ are barely observable due to their thinnes and close proximity to a boundary. 

\section{Resonances at $\mathcal{O}(\epsilon^2)$} 
 
As in the 2D case, we perform a canonical transformation in order to 
remove the perturbation at $\mathcal{O}(\epsilon)$, the associated   
 generatrix function being of the form:

\begin{equation} 
F(\bm{J},\bm{\theta})= \bm{J}\cdot\bm{\theta} +\, \epsilon \Phi(\bm{J},\bm{\theta}), 
\end{equation} 
where 

\begin{equation} 
\Phi(\bm{J},\bm{\theta})=\sum_{\bm{m}\in\mathcal{Y}}\Phi_{\bm{m}}(\bm{J})\sin (\bm{m}\cdot 
\bm{\theta})+ \sum_{\bm{m}\in\mathcal{Z}}\Phi_{\bm{m}}(\bm{J})\sin (\bm{m}\cdot \bm{\theta}), 
\end{equation} 
with $\Phi_{\bm{m}}(\bm{J})=-V_{\bm{m}}(\bm{J})/\bm{m}\cdot\bm{\omega}(\bm{J})$. 
 
The new Hamiltonian is described by the same formal expression given in equation (\ref{Hamiltoniano2}), 
where the first and second terms within braces adopt the values: 

\begin{equation} 
\frac{1}{2}\frac{\partial^2 H_0(\bm{J})}{\partial J_j \partial J_i} \frac{\partial \Phi(\bm{J},\bm{\varphi})}{\partial \varphi_j}\frac{\partial \Phi(\bm{J},\bm{\varphi})}{\partial \varphi_i}=  
\end{equation} 

\begin{align*}
    =&\sum_{\bm{m}\in\mathcal{Y}}\sum_{\bm{m'}\in\mathcal{Y}}\delta_{\bm{m},\bm{m'}}\frac{(m_1m_1'J_1^{2/3}J_2^{2/3}+m_2m_2'J_1^{4/3})}{(m_1J_1^{1/3}+m_2J_2^{1/3})(m_1'J_1^{1/3}+m_2'J_2^{1/3})} 
    \left\{\cos\big((\bm{m}+\bm{m'})\cdot\bm{\varphi}\big)+\cos\big((\bm{m}-\bm{m'})\cdot\bm{\varphi}\big)\right\}+ \\
    &+\sum_{\bm{m}\in\mathcal{Z}}\sum_{\bm{m'}\in\mathcal{Z}}\delta_{\bm{m},\bm{m'}}\frac{(m_1m_1'J_1^{2/3}J_3^{2/3}+m_3m_3'J_1^{4/3})}{(m_1J_1^{1/3}+m_3J_3^{1/3})(m_1'J_1^{1/3}+m_3'J_3^{1/3})} 
    \left\{\cos\big((\bm{m}+\bm{m'})\cdot\bm{\varphi}\big)+\cos\big((\bm{m}-\bm{m'})\cdot\bm{\varphi}\big)\right\}+ \\ 
    &+\sum_{\bm{m}\in\mathcal{Y}}\sum_{\bm{m'}\in\mathcal{Z}}\delta_{\bm{m},\bm{m'}}\frac{m_1m_1'J_1^{2/3}J_2^{1/3}J_3^{1/3}}{(m_1J_1^{1/3}+m_2J_2^{1/3})(m_1'J_1^{1/3}+m_3'J_3^{1/3})} 
    \left\{\cos\big((\bm{m}+\bm{m'})\cdot\bm{\varphi}\big)+\cos\big((\bm{m}-\bm{m'})\cdot\bm{\varphi}\big)\right\}+ \\ 
    &+\sum_{\bm{m}\in\mathcal{Z}}\sum_{\bm{m'}\in\mathcal{Y}}\delta_{\bm{m},\bm{m'}}\frac{m_1m_1'J_1^{2/3}J_2^{1/3}J_3^{1/3}}{(m_1J_1^{1/3}+m_3J_3^{1/3})(m_1'J_1^{1/3}+m_2'J_2^{1/3})} 
    \left\{\cos\big((\bm{m}+\bm{m'})\cdot\bm{\varphi}\big)+\cos\big((\bm{m}-\bm{m'})\cdot\bm{\varphi}\big)\right\},\\
\intertext{where $\delta_{\bm{m},\bm{m'}}= 2^3 3^{2/3}\beta^{20/3}\alpha_{\bm{m}}\alpha_{\bm{m'}}$ and
\begin{equation} 
\frac{\partial V(\bm{J},\bm{\varphi})}{\partial J_i}\frac{\partial \Phi(\bm{J},\bm{\varphi})}{\partial \varphi_i}= 
\end{equation}} 
  =&-\sum_{\bm{m}\in\mathcal{Y}}\sum_{\bm{m'}\in\mathcal{Y}}2\delta_{\bm{m},\bm{m'}}\frac{(2m_1'J_1^{1/3}J_2^{2/3}+m_2'J_1^{4/3}J_2^{-1/3})}{(m_1'J_1^{1/3}+m_2'J_2^{1/3})}\left\{\cos\big((\bm{m}+\bm{m'})\cdot\bm{\varphi}\big)+\cos\big((\bm{m}-\bm{m'})\cdot\bm{\varphi}\big)\right\}+\\ 
  & -\sum_{\bm{m}\in\mathcal{Z}}\sum_{\bm{m'}\in\mathcal{Z}}2\delta_{\bm{m},\bm{m'}}\frac{(2m_1'J_1^{1/3}J_3^{2/3}+m_3'J_1^{4/3}J_3^{-1/3})}{(m_1'J_1^{1/3}+m_3'J_3^{1/3})}\left\{\cos\big((\bm{m}+\bm{m'})\cdot\bm{\varphi}\big)+\cos\big((\bm{m}-\bm{m'})\cdot\bm{\varphi}\big)\right\}+ \\ 
  & -\sum_{\bm{m}\in\mathcal{Y}}\sum_{\bm{m'}\in\mathcal{Z}}2^2\delta_{\bm{m},\bm{m'}}\frac{m_1'J_1^{1/3}J_2^{1/3}J_3^{1/3}}{(m_1'J_1^{1/3}+m_3'J_3^{1/3})}\left\{\cos\big((\bm{m}+\bm{m'})\cdot\bm{\varphi}\big)+\cos\big((\bm{m}-\bm{m'})\cdot\bm{\varphi}\big)\right\}+\\ 
  & -\sum_{\bm{m}\in\mathcal{Z}}\sum_{\bm{m'}\in\mathcal{Y}}2^2\delta_{\bm{m},\bm{m'}}\frac{m_1'J_1^{1/3}J_2^{1/3}J_3^{1/3}}{(m_1'J_1^{1/3}+m_2'J_2^{1/3})}\left\{\cos\big((\bm{m}+\bm{m'})\cdot\bm{\varphi}\big)+\cos\big((\bm{m}-\bm{m'})\cdot\bm{\varphi}\big)\right\}.
\end{align*}

Therefore, the Hamiltonian may be recast as: 
\begin{equation}
  \begin{split} 
    \mathcal{H}(\bm{J},\bm{\varphi})=H_0(\bm{J})\;+\;\epsilon^2\sum_{n\in\mathcal{A}}\left\{\vphantom{\sum_{\bm{m}\in\mathcal{Z}}}\right.&\sum_{\bm{m}\in\mathcal{Y}}\sum_{\bm{m'}\in\mathcal{Y}}[a(\bm{n},\bm{m},\bm{m'},\bm{J})+a(\bm{n},\bm{m},-\bm{m'},\bm{J})]\cos(\bm{n}\cdot\bm{\varphi})+\\ 
    &+\sum_{\bm{m}\in\mathcal{Z}}\sum_{\bm{m'}\in\mathcal{Z}}[b(\bm{n},\bm{m},\bm{m'},\bm{J})+b(\bm{n},\bm{m},-\bm{m'},\bm{J})]\cos(\bm{n}\cdot\bm{\varphi})+ \\ 
    &+\sum_{\bm{m}\in\mathcal{Y}}\sum_{\bm{m'}\in\mathcal{Z}}[c(\bm{n},\bm{m},\bm{m'},\bm{J})+c(\bm{n},\bm{m},-\bm{m'},\bm{J})]\cos(\bm{n}\cdot\bm{\varphi})+ \\ 
    &\left.+\sum_{\bm{m}\in\mathcal{Z}}\sum_{\bm{m'}\in\mathcal{Y}}[d(\bm{n},\bm{m},\bm{m'},\bm{J})+d(\bm{n},\bm{m},-\bm{m'},\bm{J})]\cos(\bm{n}\cdot\bm{\varphi})\right\},
  \end{split}
\end{equation}
where $\mathcal{A}$ is the set of harmonic vectors arising through any combination of vectors from 
$\mathcal{Y}\cup\mathcal{Z}$ and whose first nonzero component is positive, and the coefficients 
$a$, $b$, $c$, and $d$ are defined as follows: 
\begin{equation*} 
  a(\bm{n},\bm{m},\bm{m'},\bm{J})=
  \begin{cases} 
    \delta_{\bm{m},\bm{m'}}\left[\frac{(m_1m_1'J_1^{2/3}J_2^{2/3}+m_2m_2'J_1^{4/3})}{(m_1J_1^{1/3}+m_2J_2^{1/3})(m_1'J_1^{1/3}+m_2'J_2^{1/3})}-2\frac{(2m_1'J_1^{1/3}J_2^{2/3}+m_2'J_1^{4/3}J_2^{-1/3})}{(m_1'J_1^{1/3}+m_2'J_2^{1/3})}\right]& \text{if }\pm \bm{n}=\bm{m}+\bm{m'},\\ 
    0 &\text{if }\pm \bm{n}\neq\bm{m}+\bm{m'}.  
  \end{cases} 
\end{equation*} 
\begin{equation*}  
  b(\bm{n},\bm{m},\bm{m'},\bm{J})=
  \begin{cases} 
    \delta_{\bm{m},\bm{m'}}\left[\frac{(m_1m_1'J_1^{2/3}J_3^{2/3}+m_3m_3'J_1^{4/3})}{(m_1
        J_1^{1/3}+m_3J_3^{1/3})(m_1'J_1^{1/3}+m_3'J_3^{1/3})}-2\frac{(2m_1'J_1^{1/3}J_3^{2/3}+m_3'J_1^{4/3}J_3^{-1/3})}{(m_1'J_1^{1/3}+m_3'J_3^{1/3})}\right]& \text{if }\pm \bm{n}=\bm{m}+\bm{m'},\\ 
    0&\text{if }\pm \bm{n}\neq\bm{m}+\bm{m'}.  
  \end{cases} 
\end{equation*} 
\begin{equation*}
  c(\bm{n},\bm{m},\bm{m'},\bm{J})=
  \begin{cases} 
    \delta_{\bm{m},\bm{m'}}\left[\frac{m_1m_1'J_1^{2/3}J_2^{1/3}J_3^{1/3}}{(m_1J_1^
        {1/3}+m_2J_2^{1/3})(m_1'J_1^{1/3}+m_3'J_3^{1/3})}-4 \frac{m_1'J_1^{1/3}J_2^{1/3}J_3^{1/3}}{(m_1'J_1^{1/3}+m_3'J_3^{1/3})}\right]& \text{if }\pm \bm{n}=\bm{m}+\bm{m'},\\ 
0&\text{if }\pm \bm{n}\neq\bm{m}+\bm{m'}.
\end{cases}  
\end{equation*}
 
\begin{equation*} 
  d(\bm{n},\bm{m},\bm{m'},\bm{J})=
  \begin{cases} 
    \delta_{\bm{m},\bm{m'}}\left[\frac{m_1m_1'J_1^{2/3}J_2^{1/3}J_3^{1/3}}{(m_1J_1^{1/3}+m_3J_3^{1/3})(m_1'J_1^{1/3}+m_2'J_2^{1/3})}-4 \frac{m_1'J_1^{1/3}J_2^{1/3}J_3^{1/3}}{(m_1'J_1^{1/3}+m_2'J_2^{1/3})}\right]&\text{if }\pm \bm{n}=\bm{m}+\bm{m'},\\ 
    0&\text{if }\pm \bm{n}\neq\bm{m}+\bm{m'}.  
  \end{cases} 
\end{equation*} 

Further, on introducing a coefficient anologous to the one in the perturbation at $\mathcal{O}(\epsilon^2)$ for the 2D model, 
\begin{equation}
  \begin{split}
    \mathbb{D}(\bm{n},\bm{J})=& \sum_{\bm{m}\in\mathcal{Y}}\sum_{\bm{m'}\in\mathcal{Y}}
    [a(\bm{n},\bm{m},\bm{m'},\bm{J})+a(\bm{n},\bm{m},-\bm{m'},\bm{J})]+\sum_{\bm{m}
      \in\mathcal{Z}}\sum_{\bm{m'}\in\mathcal{Z}}[b(\bm{n},\bm{m},\bm{m'},\bm{J})+b(\bm{n}
    ,\bm{m},-\bm{m'},\bm{J})]+\\
    &+\sum_{\bm{m}\in\mathcal{Y}}\sum_{\bm{m'}\in\mathcal{Z}}[c(\bm{n},\bm{m},\bm{m'},\bm{
      J})+c(\bm{n},\bm{m},-\bm{m'},\bm{J})]+\sum_{\bm{m}\in\mathcal{Z}}\sum_{\bm{m'}\in
      \mathcal{Y}}[d(\bm{n},\bm{m},\bm{m'},\bm{J})+d(\bm{n},\bm{m},-\bm{m'},\bm{J})],
  \end{split}
\end{equation}
the Hamiltonian allows for the expression: 
 
\begin{equation} 
\mathcal{H}(\bm{J},\bm{\varphi})=H_0(\bm{J})+\epsilon^2\sum_{\bm{n}\in\mathcal{A}} 
\mathbb{D}(\bm{n},\bm{J})\cos(\bm{n}\cdot\bm{\varphi}). 
\end{equation} 
 
Close to any resonant action associated to a resonant vector 
at $\mathcal{O}(\epsilon^2)$, $\bm{n}$, we could apply the pendulum approximation to get the same formal expression for the resonant Hamiltonian given 
by (\ref{resonantHamiltonian2}). 
 
The variations of the energy components are similar to those of $\mathcal{O}(\epsilon)$; indeed,    
\begin{displaymath} 
\Delta h_i= \frac{8}{3}AJ_i^{1/3}(\mu|\mathcal{U}_0|)^{1/2}n_i. 
\end{displaymath} 
 
For those $\bm{n}$ with either $n_3=0$ or $n_2=0$, the resonant curves are given by equations (\ref{curvaY}) and (\ref{curvaZ}) respectively, with the pertinent substitution of $m_i$ and $I_i$ by $n_i$ and $J_i$. Meanwhile, whenever $n_1=0$ the resonant curves are given by  
\begin{equation} 
\label{curvaX} 
\left\{ \begin{array}{l} 
J_2^r=\frac{1}{\left(1+(n_2/n_3)^4\right)^{3/4}}\left\{\frac{h}{A}-(J_1^r)^{4/3}\right\}^{3/4}, \\ 
J_3^r=\frac{1}{\left(1+(n_3/n_2)^4\right)^{3/4}}\left\{\frac{h}{A}-(J_1^r)^{4/3}\right\}^{3/4}, 
\end{array}\right. 
\end{equation} 
\noindent 
with $J_1^r \in [0,(\frac{h}{A})^{3/4}]$. 

Let $\mathcal{V}_r(\epsilon^2,1/23^2)$ be the set of resonant vectors which 
belong to $\mathcal{A}$ and that can be constructed by at least one pair 
$(\bm{m},\bm{m'})$ such that $\mathcal{O}(\alpha_{\bm{m}}\alpha_{\bm{m'}})\leq 
\mathcal{O}(1/23^2)$.
On computing the elements of $\mathcal{V}_r(\epsilon^2,1/23^2)$ we learn that 
this set consists of fifteen  vectors with one null component together  with forty eight having all its three components different from zero.

\begin{figure}[!ht] 
   \begin{tabular}{cc} 
   \hspace{-4mm}\includegraphics[width=.5\textwidth]{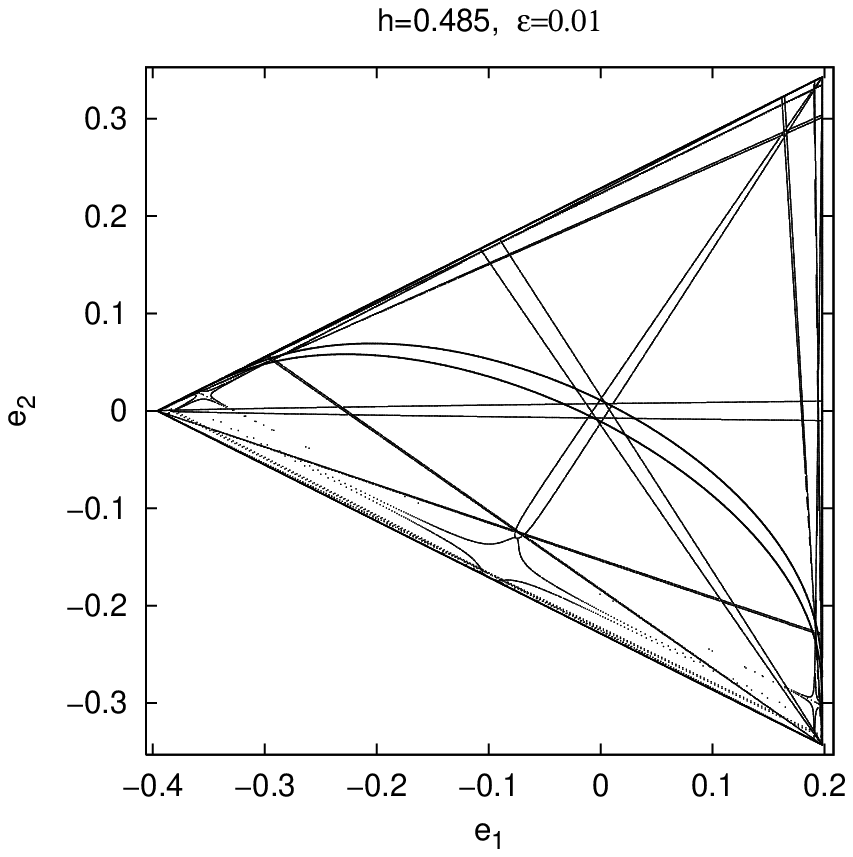} 
   \hspace{-4mm}\includegraphics[width=.5\textwidth]{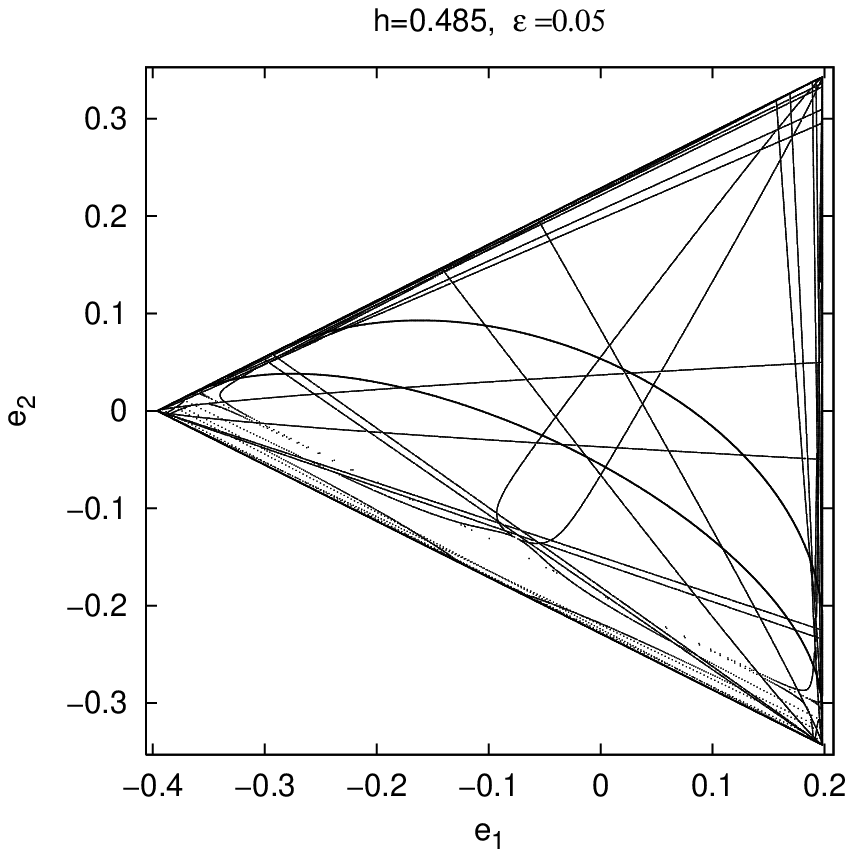} 
   \end{tabular} 
  \caption{Resonances at $\mathcal{O}(\epsilon^2,1/23^2)$ 
   for $\epsilon=0.01$ (on the left) and $\epsilon=0.05$ (on the right).}
  \label{3Dresonancias2} 
\end{figure}

Fig. \ref{3Dresonancias2} shows the area of the canonical energy surface $(h=0.485)$ occupied by those resonances at $\mathcal{O}(\epsilon^2,1/23^2)$ that have one null component together with the $(2,-1,-1)$ resonance, for the same two values of the perturbative parameter used for Fig. \ref{3Dresonancias1}. 
 
\begin{figure}[!ht]
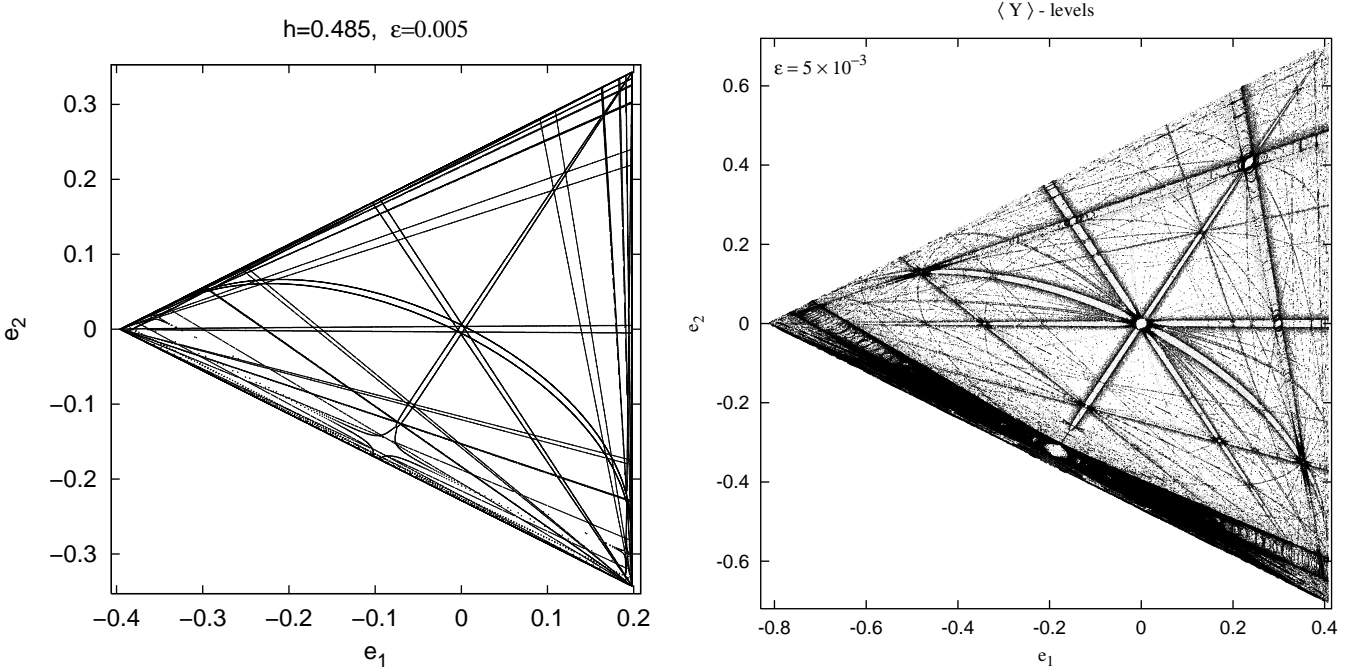
 
\begin{tabular}{cc}
   \hspace{-4mm}\includegraphics[width=.48\textwidth]{reson7c-9i.h0.485.eps0.005plusN2-1-1.reduc.epsi}
   \hspace{4mm}\includegraphics[width=.48\textwidth]{det0rot.epsi}
   \end{tabular}
  \caption{Resonances up to $\mathcal{O}(\epsilon^2,1/23^2)$ for $\epsilon=0.005$
(on the left) and the actual resonance structure obtained with the MEGNO (on the right)
for the same energy normalized to $h=1$.} 
  \label{3Dresonancias1y2} 
\end{figure}

Such a  subset of $\mathcal{V}_r(\epsilon^2,1/23^2)$ as well as  the 
complete set $\mathcal{V}_r(\epsilon,1/23^2)$ have been considered 
in Fig. \ref{3Dresonancias1y2}--left for $\epsilon=0.005$.  
This picture should be compared with the contour--plot obtained by means of the 
MEGNO for the very same value of $\epsilon$, which is displayed in 
Fig. \ref{3Dresonancias1y2}--right (taken from \cite{CGS03}). 
This numerical exploration evinces that the resonances which strongly manifest are those with just one null component (i.e. the straight ones) and the $(2,-1,-1)$ resonance (the one showing a curved shape).
 
A glance at Fig. \ref{3Dresonancias1y2}--left reveals that in some intersections
between $\mathcal{O}(\epsilon)$ and $\mathcal{O}(\epsilon^2)$ resonances, the 
widths of the latter tend asymptotically to infinity.
This is due to the emergence of small denominators in the Fourier coefficients
of the perturbation, a fact that reminds us that the canonical transformation 
perfomed in order to eliminate the perturbation terms proportional to $\epsilon$ is no  
longer valid in the neighbourhood of any $\mathcal{O}(\epsilon)$ resonance. 

A good example of this behaviour is the intersection between the $(0,1,-1)$ and
the $(2,-1,0)$ resonances. The former is an $\mathcal{O}(\epsilon^2)$ resonance
that starts on the top right--hand corner of the energy surface and gets through
the middle of it while the latter is of $\mathcal{O}(\epsilon)$ and can be 
identified as the 
widest of the resonances departing from the bottom right--hand corner 
of the energy surface. 
   
\section{Analytical estimate of the critical value of the perturbation parameter for the 3D system}

As it can be seen from Fig. \ref{3Dresonancias1y2}, many resonances have a triangular shape. Such is the case of all those resonances associated to a vector having either its second or its third component equal to zero.  

It can be demonstrated instead that 
for a resonant vector with its first component $n_1$ equal to zero,   
the coefficientes $a(\bm{n},\bm{m},\pm\bm{m'},\bm{J})$ 
and $b(\bm{n},\bm{m},\pm\bm{m'},\bm{J})$ are null  
$\forall \,(\bm{m}, \bm{m'})$.   
Thus, the contribution of such a vector proceeds only through its 
concomitant coefficients $c(\bm{n},\bm{m},\pm\bm{m'},\bm{J})$ and $d(\bm{n},\bm{m},\pm\bm{m'},\bm{J})$. 
Further, from equation (\ref{curvaX}) it can be stated that the resonance width tends to zero when $J_1$ approaches either $0$ or $(h/A)^{3/4}$. 
Consequently, the regions encompassed by the separatrices of resonances for which $n_1=0$ do not have a triangular shape. 
 
On estimating $\epsilon_{\mathrm{c}}$ for the 3D model, we are compelled to make 
somewhat strong simplifications:  
(i)~we take as $\epsilon_{\mathrm{c}}$ the value of $\epsilon$ for which the total area 
covered by resonant regions $(A_r)$ equals 90\% of the whole area of the energy surface $(A_h)$;
(ii)~we approximate by triangles the resonant regions 
corresponding to resonant vectors with one null component;
(iii)~we approximate by two triangles the resonant region corresponding to 
the $(2,-1,-1)$ resonance;
(iv)~we do not consider any further resonance; 
(v)~we add up the area of each resonance disregarding the intersections 
due to crossings of resonances, 
so that those regions corresponding to two different resonances are 
considered twice. 
 
\begin{figure}[!ht] 
\begin{tabular}{cc}
   \hspace{-4mm}\includegraphics[width=.5\textwidth]{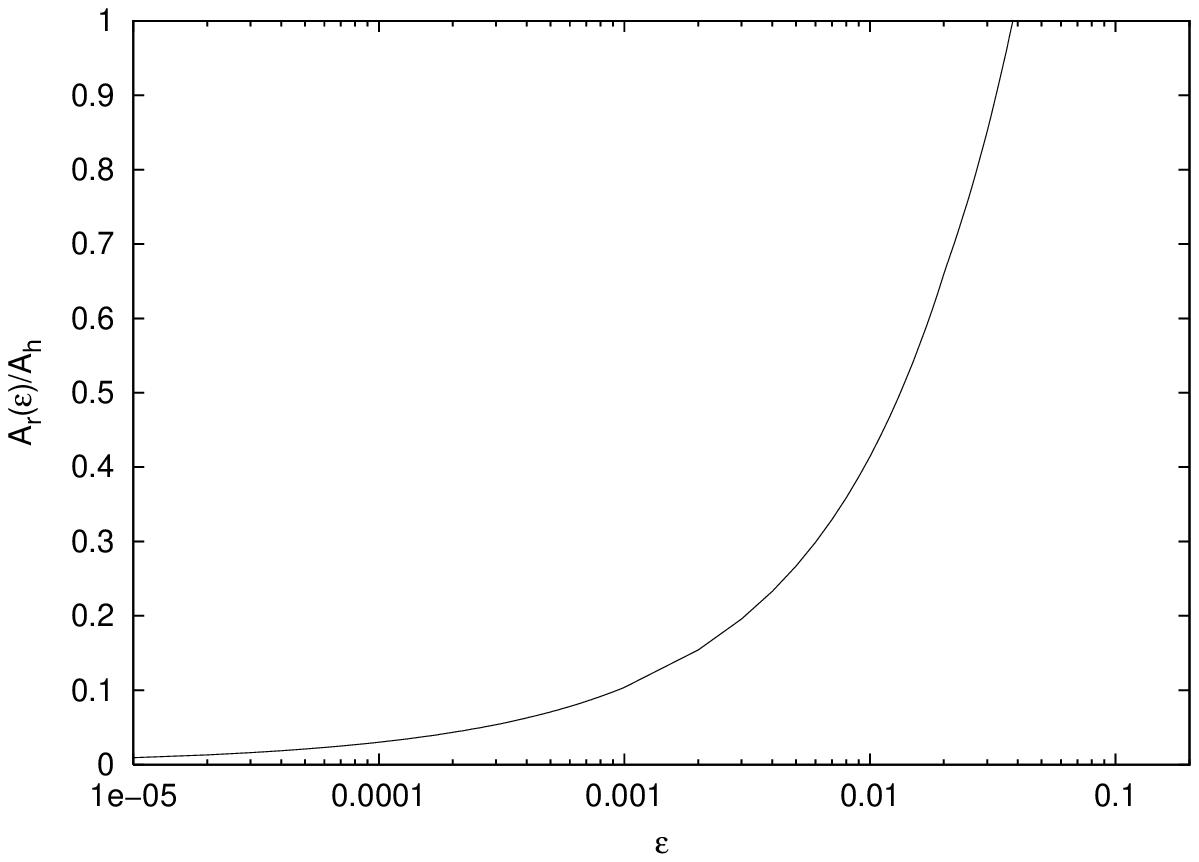}
   \hspace{2mm}\includegraphics[width=.53\textwidth]{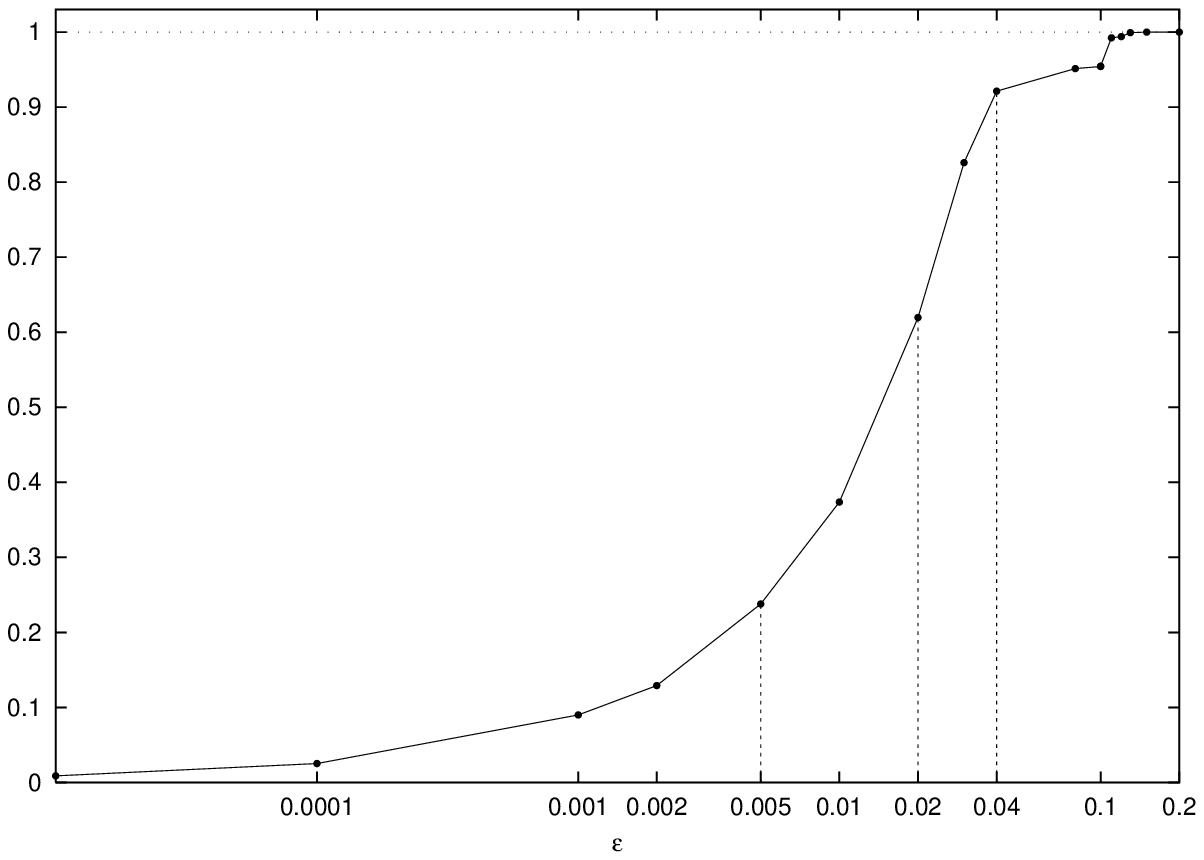}
 \end{tabular}
 \caption{ $A_r/A_h$ (on the left) and fraction of chaotic motion (on the right) both vs. the perturbation parameter, in logaritmic scale} 
 \label{3Dareatriangulos} 
\end{figure}

In Fig. \ref{3Dareatriangulos}--left we have plotted the fraction $A_r/A_h$ for the perturbation parameter varying in the range 
$\epsilon \in [0.00001,0.2]$.
There it can be observed that $A_r(\epsilon)$ reaches 90\% of 
$A_h$ for some $\epsilon_{\mathrm{c}}$ between $0.03$ and $0.04$. 
This result is in quite good agreement with that arising from 
Fig. \ref{3Dareatriangulos}--right 
(taken from \cite{GC04}) which displays the fraction of chaotic motion
according to the MEGNO values, for the same range of $\epsilon$. 
 
\section{Discussion} 

We have checked out the accuracy of the overlap criterion 
when applied to a simple near--integrable Hamiltonian system in both its 2D and 3D version. 
To this end, we have computed the unperturbed resonances up to order  
${\mathcal O}(\epsilon^2)$ for both systems, and modelled each 
resonance by means of the pendulum approximation in order to 
 estimate the theoretical critical value of the perturbation parameter 
for a global transition to chaos. 

By performing several surface of sections for the  
the 2D case we have derived an empirical value to be compared to our theoretical estimation, 
both being in good agreement. 
For the 3D case a theoretical estimate of the critical parameter has been attained, 
which is shown to match the one given in \cite{GC04}, where such a value is achieved     
on computing the fraction of chaotic motion vs. $\epsilon$ according to the MEGNO values. 
 
Let us remark that the conception of transition to global
chaos assumed for the numerical estimate of $\epsilon_{\mathrm{c}}$ in the 2D case is of a 
different nature from the one adopted for the 3D system.
Actually, the 2D system is considered to be globally chaotic if   
the chaotic component of phase space appears almost fully connected, 
while in the 3D case the system is regarded as globally chaotic when 
at most 10\% of the energy surface corresponds to invariant tori. Notice that  
in the latter case, though it is very likely that the
chaotic component be connected when resonances do overlap in a mostly chaotic 
 phase space,  
nothing could be asserted about the existence of a fully connected region of 
unstable motion (see 
\cite{GC04}, \cite{CGP06} and \cite{CG07} for a thorough discussion).

Therefore, from both theoretical and numerical results we may assert that a suitable 
estimate for the critical value of the perturbation parameter could be obtained
by means of the overlap criterion when considering resonances up to 
$\mathcal{O}(\epsilon^2)$.  
Indeed, regarding terms just up to $\mathcal{O}(\epsilon)$ largely overestimates 
$\epsilon_{\mathrm{c}}$, as already shown by \cite{CH79} for the Standard Map.

\section*{Acknowledgments} 
 
This work was partially supported by grants of the CONICET, University of La Plata an the ANPCyT (Argentina). Mestre is grateful to C. Llinares for 
his valuable advice on certain numerical issues and to F. Bareilles for his 
helpful teaching of some Fortran77 resources.

\end{document}